\documentclass{article}
\usepackage{spconf,amsmath,graphicx,hyperref}

\usepackage{float}

\usepackage{subcaption,graphicx}
\graphicspath{ {./img/} }

\usepackage{amsmath}
\usepackage{mathtools}
\usepackage{amssymb}
\usepackage{outline}
\usepackage{booktabs}

\usepackage{orcidlink}

\usepackage{bm}      % for \bm macro

\newcommand{\figref}[1]{Figure~\ref{fig:#1}}
\newcommand{\tblref}[1]{Table~\ref{table:#1}}

\newcommand{\B}[1]{\textbf{#1}}

\usepackage{url}
\usepackage{enumitem}

\title{MORPHO-VITS: VARIATIONAL INFERENCE WITH MORPHOLOGICAL MODELING FOR END-TO-END SPEECH SYNTHESIS OF A TONAL BANTU LANGUAGE}
\name{Antoine Nzeyimana\,\orcidlink{0000-0003-4567-3471}}
\address{University of Massachusetts Amherst, USA}
\begin{document}
%\ninept
%
\maketitle

\begin{abstract}
Text-to-speech models for Bantu tonal languages are challenged by a tonal system that is rooted in both the lexis (i.e., the inventory of words, stems, and affixes) and the grammar (i.e., morpho-syntax). To complicate matters, the standard writing systems of these languages often omit tone markings and syllable duration information, which must be disambiguated by the reader based on context. Motivated by linguistic descriptions of Bantu language tone systems, we propose an end-to-end text-to-speech model that augments the text encoding mechanism with a morpho-syntactic prior. We replace the standard phoneme encoder in the VITS architecture with a morpheme sequence encoder and a phoneme-to-morpheme attention network. We posit that, by using this explicit morphological modeling, we can capture the information required to produce the correct tone. Experiments conducted on the Kinyarwanda language, a tonal and morphologically complex Bantu language, reveal substantial TTS improvement from this morphological modeling. Specifically, the proposed method significantly improves the naturalness, intonation, and intelligibility of the produced synthetic voices.
\end{abstract}
\begin{keywords}
Text-to-speech, Bantu tone, Morphological modeling
\end{keywords}

\section{Introduction}
\label{sec:intro}

Text-to-speech (TTS) synthesis represents an important technology that benefits digital users in both developed and developing countries. TTS enables hands-free operation of smart devices in situations where devices lack other interactive interfaces or when the user is constrained from using their hands. TTS also empowers automated interactive voice response (IVR) systems that make information accessible to any cellular device, including feature phones, which are still common in developing countries. Voice assistance interfaces can also use TTS technology to allow people with spoken communication disabilities to communicate with others. Even though such applications are highly useful, the quality of synthetic voices for many of the world's languages remains low~\cite{rutunda2023kinyarwanda}.

Despite the availability of TTS datasets and the relative ease of creating new ones, as well as the recent availability of end-to-end TTS training frameworks, the quality of synthetic voices for tonal Bantu languages is still low, mostly due to linguistic challenges. This primarily stems from the absence of tone markings and syllable duration information in standard orthography~\cite{hurskainen2009enriching}, which necessitates implicit disambiguation during reading. In fact, producing correct tones is challenging not only for TTS systems but also for non-native speakers and language learners. This leads to a situation where TTS systems deployed in real applications are typically perceived as sounding like non-native speakers.

In the case of Kinyarwanda, a tonal and morphologically complex Bantu language spoken by $\sim${15} million people in Eastern and Central Africa, previous studies in its tonology~\cite{goldsmith2011rhythm,Kimenyi2002ATG,myers2003f0} indicate that tones are produced through both lexical and grammatical processes and are auto-segmental in nature~\cite{clements1984autosegmental,rawski2020multi}, which makes it challenging to master for both language learners and trained TTS models. In fact, there is a two-way relationship between tone and morphology, where (1) knowing the tone can improve morphological disambiguation~\cite{muhirwe2009morphological} and (2) knowing the morphology can help anticipate tone~\cite{hurskainen2009enriching} through rule-based symbolic processing. Therefore, in this work, we examine whether explicit morphological modeling can improve end-to-end TTS models for tonal Bantu languages. We posit that using a morphological analyzer and explicit morphological representation in TTS sequence models (e.g, morpheme sequence in conjunction with phoneme sequence) can capture both lexical and grammatical (i.e, morpho-syntax) patterns needed to produce correct tones.

We build upon the VITS2 model architecture~\cite{kim2021conditional,kong2023vits2}, which uses variational inference and adversarial training to learn a generative model of speech given text. We enhance the original VITS text encoder with a morpheme sequence encoder and a phoneme-to-morpheme sequence cross-encoder, through which each phoneme attends to the hidden representations of morphemes in the same word. Experiments conducted on a multi-speaker Kinyarwanda TTS corpus indicate that significant TTS quality improvement can be achieved through morphological modeling, leading to synthetic voices that are more natural and intelligible.

% \section{Related Work}
% \label{sec:related}

There are several previous studies that have attempted to enhance end-to-end TTS models with additional linguistic information. For example, the work in ~\cite{liu2024text} uses pre-trained language model embeddings and concatenates them with the text encoder hidden representations to enhance a TTS model for the Mongolian language. Similarly,~\cite{JIANG2024128430} attempts to augment Fastspeech 2~\cite{ren2020fastspeech} architecture with semantic dependency parsing and a forced tone sequence to improve a Chinese language TTS model. Graphspeech~\cite{liu2021graphspeech} proposes a graph neural network (GNN) to represent syntactic information for the English language and uses graph-to-sequence modeling to enhance English TTS model. These related works demonstrate the importance of augmenting phoneme-based text encoding with linguistic context to enhance TTS modeling. However, these methods do not address the challenge of modeling intonation in Bantu tonal languages. Our proposed method is motivated by previous findings in Bantu language phonology that describe the tone system. The fact that the linguistic lexicon (e.g., words, stems, and affixes), morphology, and syntax all contribute to the realization of tone motivated us to conduct this empirical study using morpheme sequence modeling as a way of capturing the underlying prior for tone production.

\section{Proposed Methods}
\label{sec:methods}

\subsection{Architecture overview}

\begin{figure}[!ht]
 \centering
 \includegraphics[width=0.95\linewidth]{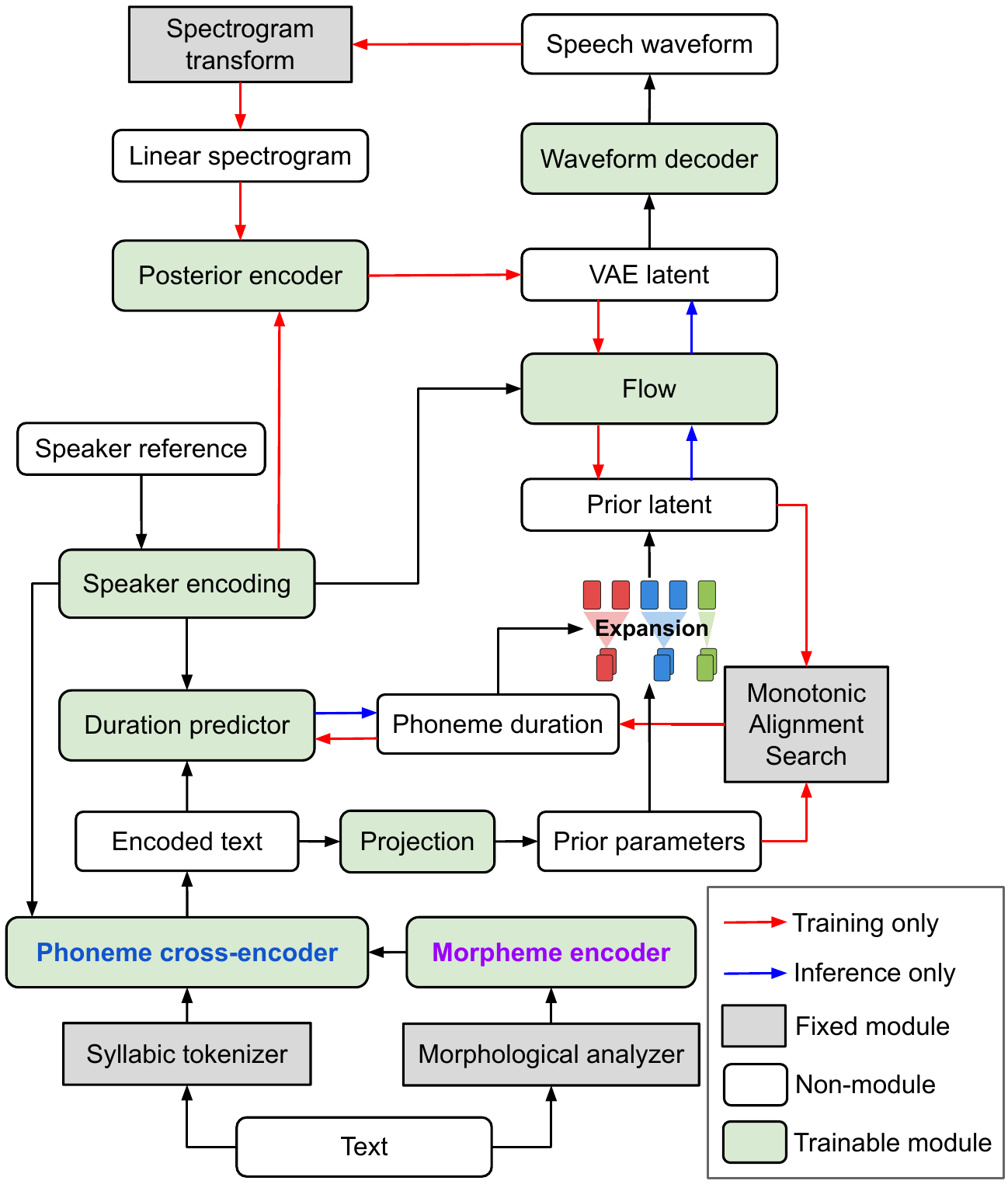}
 \caption{The overall architecture is based on he VITS framework~\cite{kim2021conditional,kong2023vits2}. We enrich the text prior encoder with a morpheme encoder and a phoneme-to-morpheme cross-encoder to capture detailed morpho-syntactic patterns in the input text which are needed to disambiguate intonation.}
\label{fig:arch}
\vspace{-0.2in}
\end{figure}

Our overall model architecture is captured in~\figref{arch} and is based on the VITS architectural framework~\cite{kim2021conditional,kong2023vits2}. Specifically, the VITS architecture uses a conditional variational auto-encoder (CVAE) to approximate the intractable marginal log-likelihood $\log p_{\theta}(x|c)$ of the speech waveform $x$ given the text $c$ by optimizing the evidence lower-bound or ELBO, i.e.:
\begin{align}
    &\log p_{\theta}(x|c) = \log \int p_{\theta}(x,z|c)dz \notag\\
    &\geq \int q_{\phi}(z|x)\log\frac{p_{\theta}(x,z|c)}{q_{\phi}(z|x)} dz \notag\\
    &= \int q_{\phi}(z|x)\log\frac{p_{\theta}(x|z)p(z|c)}{q_{\phi}(z|x)} dz \notag\\
    &= E_{q_{\phi}(z|x)}[\log p_{\theta}(x|z)] - D_{KL}(q_{\phi}(z|x)||p_{\theta}(z|c)), \label{eq:elbo}
\end{align}
where $z$ is the VAE latent variable capturing learned acoustic features, and $D_{KL}$ represents the Kullback-Leibler (KL) divergence between the prior  $p_{\theta}(z|c)$ and posterior $q_{\phi}(z|x)$.

We use a lightweight, high-fidelity waveform decoder from ~\cite{kawamura2023lightweight}, which performs multi-band generation of spectrograms from acoustic features $z$ and inverse short time Fourier transform (iSTFT) to synthesize speech waveforms.

In a typical VITS model architecture, the prior encoder learns prior parameters from text by using a bidirectional transformer encoder over the phoneme sequence. Those prior parameters are further expanded to a frame-level representation by using monotonic alignment search (MAS)~\cite{kim2020glow}, and a flow-based reversible network~\cite{lipman2022flow} is used to match the latent acoustic features more efficiently. Additional modules include the duration predictor, which uses expansion information from the MAS module to learn each phoneme duration multiplier from the input text. The duration predictor is then used during inference to expand prior parameters to the acoustic frame-level representation. Similar to other works in multi-speaker TTS, we also use a speaker encoder to condition the network on speaker encodings, which can be mapped from a speaker identifier through embedding or encoded from sample speaker audio data for voice cloning applications.

\subsection{Enriching text prior with morphology}

Our main contribution to the Morpho-VITS architecture is to enrich the prior encoder with morpho-syntactic information. Specifically, instead of encoding text using a standard phoneme encoder, we use two modules: one for encoding the morpheme sequence and another for the phoneme-to-morpheme cross encoder. For Kinyarwanda, which, like other Bantu languages, typically uses open-syllables, we tokenize the input text using a syllabic tokenizer to produce vowels and consonant clusters that we use for phoneme embedding. Previous work~\cite{nzeyimana2023kinspeak} has found this type of tokenization to be more effective for speech recognition. In parallel to syllabic tokenization, we also use a morphological analyzer to segment words into morpheme sequences, which attempts to capture the morphological structure of words.

\begin{figure}[!ht]
 \centering
 \includegraphics[width=0.7\linewidth]{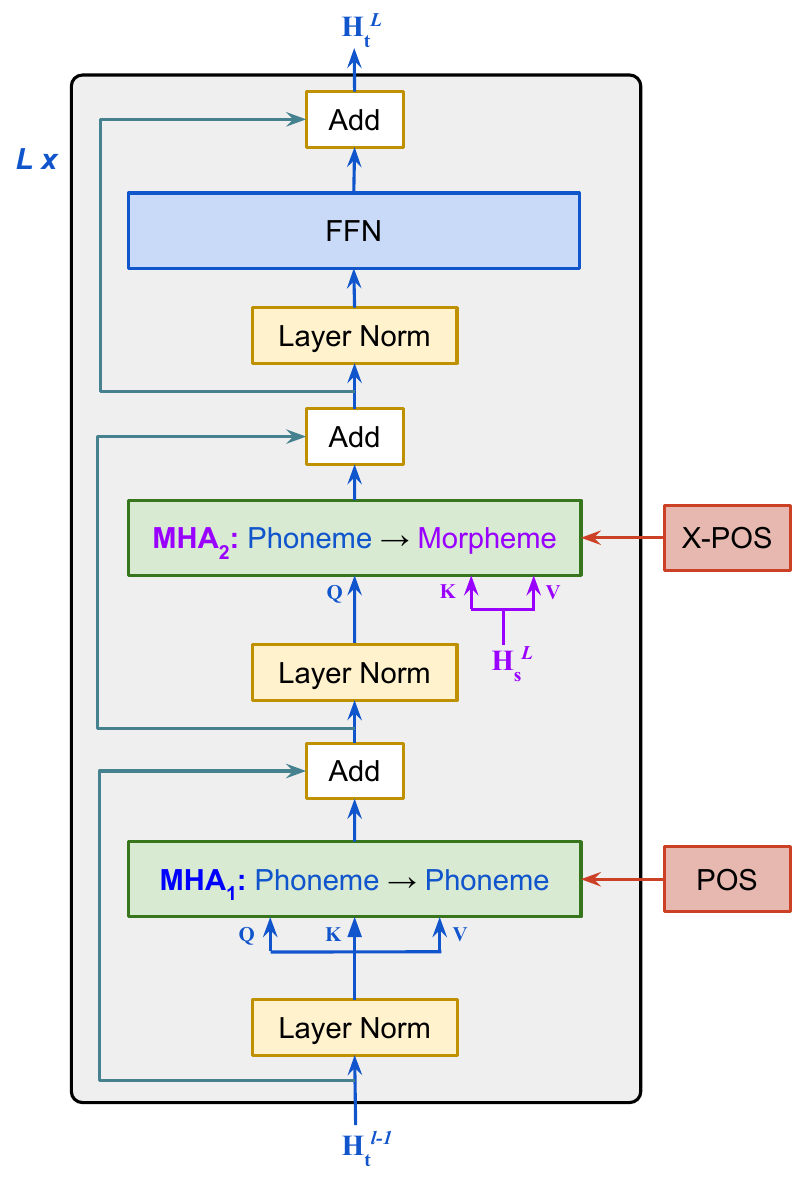}
 \caption{Our phoneme-to-morpheme cross-encoder layer uses two attention layers, one for phoneme self-attention and the other for phoneme-to-morpheme cross-attention. We use relative positional embeddings as bias terms to the logits of the attention layers.}
\label{fig:cross-encoder}
\vspace{-0.1in}
\end{figure}

The architecture of our phoneme-to-morpheme cross-encoder layer is shown in~\figref{cross-encoder}. This architecture is inspired by the Transformer decoder architecture for machine translation, except that our cross-encoder is bidirectional. Specifically, we use two attention modules: one for target-to-target self-attention and the other for target-to-source attention. In this setup, the source is the morpheme sequence, while the target is the phoneme sequence.

To give a concrete example, consider the two conjugations of the verb ``kuririmba''(to sing): (1) ``\B{bazaririmba}'' (\textit{they will sing}) and (2) ``\B{uwaririmbye}'' (\textit{the one who sang}).
The syllable-based tokenization will give the following phoneme sequences:
\vspace{-0.1in}
\begin{align*}
    &1.~ \texttt{b-a-z-a-r-i-r-i-mb-a} \\
    &2.~ \texttt{u-w-a-r-i-r-i-mby-e}
\end{align*}
while their morphological segmentation gives the following morpheme sequences:
\begin{align*}
    &1.~ \texttt{V:2:ba-V:4:zaa-V:10:rírimb-V:18:a} \\
    &2.~ \texttt{V:0:u-V:2:u-V:4:á-V:10:ríriimb-V:18:ye}
\end{align*} with ``V'' indicating that they are verbal forms and the numbers indicating morpheme slots (See ~\cite{nzeyimana2020morphological,goldsmith2011rhythm} for Kinyarwanda verb morphology).
Note how the actual high tones are caused by the tone-bearing past tense morpheme \texttt{V:4:á} and the tone-bearing stem \texttt{V:10:ríriimb} , while the future tense morpheme \texttt{V:4:zaa} has no high tone. However, in some syntactic contexts, the future tense morpheme \texttt{zaa} can also have a high tone. For example, in the phrase ``\B{Ni bo bazaririmba}'' (\textit{they are the ones who will sing}), the verb will have tone markings as: \texttt{bázáaríriimba}. Therefore, both morphology and syntax have an effect on tone realization.

Given a ``target'' sequence of phoneme hidden states $H_t = (t_1, t_2, ... t_n)$ and the corresponding ``source'' sequence of morpheme hidden states $H_s = (s_1, s_2, ... s_n)$, the cross-attention (MHA$_2$ module in~\figref{cross-encoder}) logits $\alpha_{ij}$ between the phoneme at target position $i$ and the morpheme at source position $j$ are computed as:
\begin{equation}
  \begin{aligned}
    \alpha_{ij} = 
        \begin{cases} 
            \frac{1}{\sqrt{d_k}} (t_iW_Q)(s_jW_K)^T + r_{j-i}, & \text{if } stk(i,j)\\
            -\infty,   & \text{otherwise}
        \end{cases}
  \end{aligned}
\label{eq:logits}
\end{equation}
where $W_Q$ and $W_K$ are learnable projection matrices, $r_{j-i}$ is the relative cross-positional embedding (X-POS module in~\figref{cross-encoder}) and $stk(i,j)$ function indicates whether the phoneme at target position $i$ and the morpheme at source position $j$ belong to the same token/word.
The output of the cross-attention layer is thus:
\begin{equation}
t^{'}_i=\sum_{j=1}^n \frac{\exp (\alpha_{ij})}{\sum_{j'=1}^n\exp (\alpha_{ij'})}(s_jW_{V}), \\
\label{eq:output}
\end{equation}
with $W_V$ being a learnable projection matrix.
The phoneme-to-phoneme self-attention module (MHA$_1$ in~\figref{cross-encoder}) uses the standard scaled dot-product attention with a relative position bias. The morpheme encoder uses the standard transformer encoder architecture, also with scaled dot-product attention and relative positional bias.

\section{Experiments}
\label{sec:exp}

\subsection{Dataset}

We train two multi-speaker Kinyarwanda TTS models using datasets from three sources: (1) 6 hours of news reading speech with one speaker from the Mbaza NLP project\footnote{\url{https://huggingface.co/datasets/mbazaNLP/kinyarwanda-tts-dataset}}, 20 hours of speech data with 4 speakers from an agricultural TTS dataset\footnote{\url{https://huggingface.co/datasets/C4IR-RW/kinya-ag-tts}} , and 38.5 hours of speech data from our previous speech-text alignment project~\cite{nzeyimana2023kinspeak}. 
% The last 38.5-hours dataset is clustered into 13 pseudo-speaker ids based on cosine similarity using a pretrained speaker encoding model from~\cite{jung2022pushing}.

\subsection{Morphological analysis}

We run our experiments with a new Bantu language morphological analyzer configured for the Kinyarwanda language. The analyzer is based on two-level morphology principles~\cite{koskenniemi1983two} with a custom implementation in the Rust programming language. Similar to ~\cite{nzeyimana2020morphological}, the analyzer uses stemming data to disambiguate verbal forms, which are the most complex in Bantu languages. The analyzer also uses a Hidden Markov Model for POS tagging with bi-directional decoding, as described in~\cite{nzeyimana2022kinyabert}. The analyzer achieves 85\% accuracy on the AfriSUD~\cite{buzaaba2026afrisud} POS tagging benchmark.

\subsection{Model details}

We train a baseline VITS model with a standard phoneme encoder and our proposed Morpho-VITS architecture in~\figref{arch} using the same dataset. The models are implemented in PyTorch version 2.7.1. Both models use a phoneme hidden dimension of 192, with 8 phoneme encoder/cross-encoder layers, with speaker conditioning at the third layer, as done in~\cite{kong2023vits2}. The morpheme encoder in Morpho-VITS has 8 encoder layers with 768 morpheme hidden dimensions. The baseline VITS model has 97 million parameters, while Morpho-VITS uses 268 million parameters, with 169 million dedicated to the morpheme encoder. We train each model for 1 million gradient update steps, with a batch size of 32 and a peak learning rate of 1e-4. We train each model using 4 Nvidia RTX 5090 GPUs with distributed data parallelism, and the training takes 96 hours for the VITS baseline and 149 hours for Morpho-VITS.

\subsection{Evaluation}

We evaluate the intelligibility of synthetic voices produced by the two experimental models using automatic speech recognition (ASR). We collect a set of 10,000 Kinyarwanda sentences from a news corpus, each containing 6 to 12 words. We omit sentences that contain names of people, as these can have foreign-language (e.g., Christian names) and non-standard spelling that typically require a pronunciation dictionary to convert to Kinyarwanda phonemes.

\begin{table}[!ht]
\caption{ASR results for intelligibility evaluation, comparing VIS Baseline against our proposed Morpho-VITS model.}
    \centering
\resizebox{0.9 \linewidth}{!}{
{\renewcommand{\arraystretch}{1.2}% for the vertical padding
\begin{tabular}{l c c c}
\toprule
\B{Model}  & \B{\# samples} &  \B{CER [\%]}  &  \B{WER [\%]} \\
\midrule
VITS Baseline  &  10K   &  3.0 & 13.4 \\
Morpho-VITS (Ours)  &  10K   &  \B{1.7} & \B{9.4} \\
~ ~ w/ unseen morphemes & 900 & 2.9 & 19.3\\
~ ~ w/o unseen morphemes  &  9.1K & 1.5 & 8.3 \\
\bottomrule
\end{tabular}
    }
 }
\label{table:asr_results}
\end{table}

% ASR RESULTS:
% nm_data: 10000
% wm_data: 10000
% ref_data: 430547
% nm_data: 10000
% wm_data: 10000
% ==> Final Non-Morpho CER/WER [%] (n = 10,000): 3.01/13.40
% ==> Final Morpho-VITS CER/WER [%] (n = 10,000): 1.65/9.36
% ==> Final PARTIAL MORPHEMES Morpho-VITS CER/WER [%] (n = 900 | 9.00 %): 2.92/19.28
% ==> Final FULL MORPHEMES Morpho-VITS CER/WER [%] (n = 9,100 | 91.00 %): 1.52/8.33

Using a state-of-the-art Kinyarwanda speech recognition model~\cite{nzeyimana2023kinspeak}, the ASR results are shown in~\tblref{asr_results}. While both models achieve error rates smaller than standard noisy benchmarks like Mozilla Common Voice~\cite{ardila2020common}, Morpho-VITS still produces more intelligible speech than the VITS baseline. Considering that some of the evaluation text (9\%) contains morphemes not seen during training, we differentiate between cases where there are unseen morphemes in the text and when all morphemes were seen during training. Clearly, Morpho-VITS performance degrades with unseen morphemes.

\begin{table}[!ht]
\caption{MOS (± 95\% CI) results for naturalness and tone, and percentage of perceived intelligible clips. 2,100 audio clip pair ratings were provided by 21 listeners.}
    \centering
\resizebox{\linewidth}{!}{
{\renewcommand{\arraystretch}{1.2}% for the vertical padding
\begin{tabular}{l c c c}
\toprule
\B{Model}  &  \B{Naturalness}  &  \B{Intonation}  &  \B{\%Intelligible} \\
\midrule
VITS Baseline  &  3.33 (±0.05) & 2.95 (±0.05) & 92.1\\%\% \\
Morpho-VITS (Ours)  &  \B{3.55 (±0.05)} & \B{3.12 (±0.05)} & \B{96.7}\\%\%} \\
~ ~ w/ unseen morphemes & 3.44 (±0.18) &  3.00 (±0.16) & 93.8\\%\% \\
~ ~ w/o unseen morphemes & 3.56 (±0.06) &  3.13 (±0.05) & 97.0\\%\% \\
\bottomrule
\end{tabular}
    }
 }
\label{table:mos_results}
\end{table}

% Overall results:

% Non-Morph Naturalness MOS ==> Mean = 3.33 (±0.05) [3.2806, 3.3876] (95% CI) ~>> 333.4% , n=2089 
% Morpho-VITS Naturalness MOS ==> Mean = 3.55 (±0.05) [3.4945, 3.5998] (95% CI) ~>> 354.7% , n=2089 
% Morpho-VITS PARTIAL Morpheme's Naturalness MOS ==> Mean = 3.44 (±0.18) [3.2589, 3.6132] (95% CI) ~>> 343.6% , n=211 
% Morpho-VITS FULL Morpheme's Naturalness MOS ==> Mean = 3.56 (±0.06) [3.5046, 3.6147] (95% CI) ~>> 356.0% , n=1878 

% Non-Morph Tones MOS ==> Mean = 2.95 (±0.05) [2.9005, 2.9980] (95% CI) ~>> 294.9% , n=2089 
% Morpho-VITS Tones MOS ==> Mean = 3.12 (±0.05) [3.0666, 3.1642] (95% CI) ~>> 311.5% , n=2089 
% Morpho-VITS PARTIAL Morpheme's Tones MOS ==> Mean = 3.00 (±0.16) [2.8333, 3.1572] (95% CI) ~>> 299.5% , n=211 
% Morpho-VITS FULL Morpheme's Tones MOS ==> Mean = 3.13 (±0.05) [3.0777, 3.1800] (95% CI) ~>> 312.9% , n=1878 

% Non-Morph Intelligibility MOS ==> Mean = 0.92 (±0.01) [0.9089, 0.9321] (95% CI) ~>> 92.1% , n=2089 
% Morpho-VITS Intelligibility MOS ==> Mean = 0.97 (±0.01) [0.9593, 0.9746] (95% CI) ~>> 96.7% , n=2089 
% Morpho-VITS PARTIAL Morpheme's Intelligibility MOS ==> Mean = 0.94 (±0.03) [0.9057, 0.9711] (95% CI) ~>> 93.8% , n=211 
% Morpho-VITS FULL Morpheme's Intelligibility MOS ==> Mean = 0.97 (±0.01) [0.9625, 0.9779] (95% CI) ~>> 97.0% , n=1878 

% Naturalness/Tones Pearson R ==> 0.77

% Naturalness/Tones *Spearman R ==> 0.76

% Naturalness agreement Krippendorff’s Alpha ==> Mean = 0.52 (±0.10) [0.4140, 0.6232] (95% CI) ~>> 51.9% , n=14 

% Tones agreement Krippendorff’s Alpha ==> Mean = 0.67 (±0.06) [0.6122, 0.7334] (95% CI) ~>> 67.3% , n=14 

We also run listener evaluations to assess the quality (naturalness, intonation, and intelligibility) of synthetic voices produced by the two models. Twenty-one Kinyarwanda native speakers, each with at least a high school education, were recruited and trained on how to assign naturalness and intonation scores (1 to 5) and intelligibility labels (i.e., Yes/No). Each listener rated 100 different audio clip pairs, which were paired randomly (i.e., the listener couldn't anticipate which system generated which audio clip), and each audio clip pair was rated by two different listeners. The mean opinion score (MOS) and the percentage of intelligible clips are shown in~\tblref{mos_results}. Overall, Morpho-VITS achieves significantly better performance than the VIST baseline in terms of both naturalness, intonation, and intelligibility.

\section{Conclusions}
\label{sec:conclusions}

Motivated by linguistic descriptions of the Bantu tone systems, this work examines whether using explicit morphological modeling can improve end-to-end text-to-speech for a Bantu tonal and morphologically complex language. The proposed Morpho-VITS architecture, which uses a morpheme sequence encoder and a phoneme-to-morpheme cross encoder for text encoding, achieves significantly better performance than a baseline VITS model. Morpho-VITS performance can degrade if there are morphemes unseen in the training dataset, suggesting that its training data should be morphologically diverse and cover the vast majority of the morpheme lexicon. Future work will investigate whether the text encoder network can benefit from self-supervised pre-training.

\section{Acknowledgements}
\label{sec:ack}

This work was supported by KINLP R\&D Ltd, Rwanda.

\bibliographystyle{IEEEbib}
\bibliography{refs}

\end{document}